\documentclass[aps, pra, twocolumn, showpacs]{revtex4-1}
\usepackage{amsmath}
\usepackage{amsthm}
\usepackage{amssymb}
\usepackage{amsbsy}
\usepackage{amsfonts}
\usepackage{bm}
\usepackage{color}
\usepackage{dcolumn}
\usepackage{dsfont}
\usepackage{epsfig}
\usepackage{graphicx}
\usepackage{psfrag}

\newcommand{\beq}{\begin{eqnarray}}
\newcommand{\eeq}{\end{eqnarray}}

\newcommand{\be}{\begin{equation}}
\newcommand{\ee}{\end{equation}}
\newcommand{\bq}{\begin{eqnarray}}
\newcommand{\eq}{\end{eqnarray}}

\begin{document}
\title{Incoherent dynamics in the toric code subject to disorder}
\author{Beat R\"othlisberger$^1$, James R. Wootton$^{1,2}$, Robert M. Heath$^2$, Jiannis K. Pachos$^2$, and Daniel Loss$^1$}
\affiliation{$^1$Department of Physics, University of Basel, Klingelbergstrasse 82, CH-4056 Basel, Switzerland \\
$^2$School of Physics and Astronomy, University of Leeds, Leeds LS2 9JT, U.K.}
\date{\today}
\begin{abstract}

We numerically study the effects of two forms of quenched disorder on the anyons of the toric code. Firstly, a new class of codes based on random lattices of stabilizer operators is presented, and shown to be superior to the standard square lattice toric code for certain forms of biased noise. It is further argued that these codes are close to optimal, in that they tightly reach the upper bound of error thresholds beyond which no correctable CSS codes can exist. Additionally, we study the classical motion of anyons in toric codes with randomly distributed onsite potentials. In the presence of repulsive long-range interaction between the anyons, a surprising increase with disorder strength of the lifetime of encoded states is reported and explained by an entirely incoherent mechanism. Finally, the coherent transport of the anyons in the presence of both forms of disorder is investigated, and a significant suppression of the anyon motion is found.
\end{abstract}


\maketitle

\section{Introduction}

A working quantum computer performing meaningful calculations unarguably requires information processing to be carried out in a fault-tolerant manner \cite{Nielsen2000a, Mermin2007}. This not only means protecting the information from the action of imperfect gates, but also storing it in a reliable way during the course of computation. In the theory of quantum error correction, the state of a logical qubit can be encoded in the code space of a number of physical qubits \cite{Gottesman2010}. The resulting redundancy allows one to implement fault-tolerant quantum gates and to periodically check for the occurrence of single-qubit errors using syndrome measurements. However, this kind of active error monitoring imposes an additional overhead on an already deeply involved vision. Therefore, the idea to manipulate and store quantum states in systems that already provide `built-in' protection from errors has gained a lot of attention recently \cite{Kitaev2003, Dennis2002, Bacon2006, Nussinov2009}. A promising approach in this direction is to encode information in the degenerate topologically ordered ground state of a suitable many-body Hamiltonian. Information is encoded in an entangled state distributed across a large number of qubits and can only be distinguished and modified non-locally. In this context, Kitaev's toric code \cite{Kitaev2003} is arguably the best studied model to date. It is robust against local perturbation at zero temperature, as well as against thermal errors if long-range interaction between its fundamental excitations is present \cite{Hamma2009, Chesi2010}. 

Recent studies have focussed on coherent phenomena in the toric code that arise due to the additional presence of various forms of quenched disorder \cite{Tsomokos2011, Wootton2011, Stark2011, Bravyi2011}. Conversely, this work is primarily concerned with a numerical study of incoherent (classical) effects caused by two particular forms of randomness. First, we consider a class of models similar to the toric code, but differing from the latter in that the syndrome check operators are chosen randomly from a set of 3-body and 6-body operators. These correspond to a generalization of the (square lattice) toric code to randomized lattices. We find that these models have an advantage over the toric code for biased noise, where bit-flip and phase-flip errors occur with different probabilities. We also present strong evidence that these codes are almost optimal, in the sense that they reach error thresholds close to the overall upper bound valid for any Calderbank-Shore-Steane (CSS) code \cite{Calderbank1996, Steane1996a}. Second, we investigate the effect of random onsite potentials on the lifetime of states encoded in the regular toric code coupled to a thermal bath. We identify and describe an interesting regime, where, in the presence of long-range interactions, the lifetime of this quantum memory is enhanced for increasing disorder strength. Finally the effects of the random lattices on coherent anyon transport is investigated, both with and without additional randomness in the onsite potentials. The resultant slowdown of the anyonic motion is determined and its effect on the stability of the quantum memory is discussed.

The paper is organized as follows: Section \ref{sec:toric code} reviews the toric code, which is the basis of all further studies in this work. We then show in Sec.~\ref{sec:classical dynamics} how to simulate the classical dynamics of excitations in the systems considered subsequently and also give some details on the numerics. Our main results for the random lattice models and random onsite potentials are presented in Sections \ref{sec:random lattices} and \ref{sec:random potentials}, respectively. Then, Sec. \ref{sec:quantum dynamics} studies coherent anyon transport on the random lattices. A conclusion of our work is given in Sec.~\ref{sec:conclusions}.

\section{Review of the 2D toric code}\label{sec:toric code}

\begin{figure}[!t]
\centering
\includegraphics[width=0.9\columnwidth]{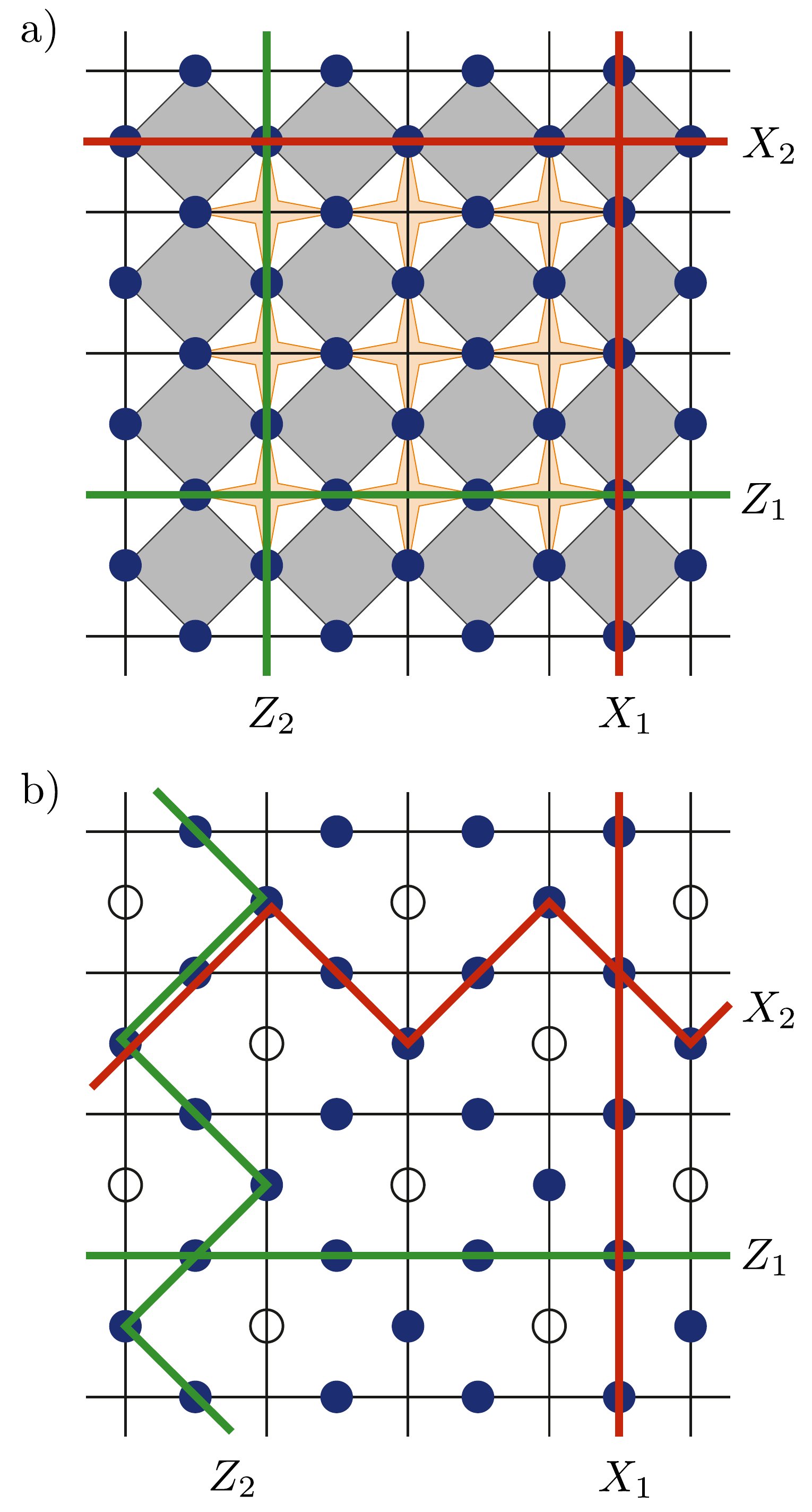}
\caption{(color online) Toric codes. a) Kitaev's original 2D toric code. Shown is a $4\times 4$ subregion of the $L \times L$ lattice. The blue solid dots on the edges of the lattice represent spins, the 4-body
plaquette and star operators are shown in light green and orange, respectively (note that stabilizers containing spins outside the figure are not shown). The four thick horizontal and vertical lines represent the logical Pauli operator strings $X_{1, 2}$ (red) and $Z_{1, 2}$ (green). b) The same region after introducing a regular pattern of spin defects. The modified plaquette and star operators are not shown yet, see Fig.~\ref{fig:random_modification}. The empty circles indicate the defect positions, i.e., the edges of the lattice where spins are removed. This requires altering the logical operators $Z_2$ and $X_2$ in the way shown. Note that all commutation relations between the logical operators are preserved.}\label{fig:toric_codes}
\end{figure}

The starting point of our investigation is Kitaev's 2D toric code \cite{Kitaev2003} which
will be modified in the following sections to incorporate randomness. We provide
here a brief outline of the original model for the sake of completeness. The 2D
toric code consists of $2L^2$ spins-$\frac{1}{2}$ with each spin placed on
an edge of an underlying $L \times L$ square lattice with periodic boundary
conditions. One then defines two sets of mutually commuting four-body operators,
called plaquettes and stars, respectively, in the following way (cf. Fig.~\ref{fig:toric_codes}a). A
plaquette is the product of the Pauli $\sigma_z$ operators associated with the
four spins belonging to a single face of the square lattice, whereas a star is
the product of the four $\sigma_x$ operators of the spins on edges adjacent to a
single vertex of the lattice. In this way, one obtains two sets of $L^2$
plaquette and star operators, out of which $L^2 -1$ in each set are independent.
Note that these operators can only have eigenvalues $+1$ and $-1$.

One can then define a subspace $\mathcal{C}$ of the total Hilbert space given by
the $2^{[2L^2 - 2(L^2 - 1)]} = 4$ states which are simultaneous eigenstates of
all independent plaquettes and stars with eigenvalue +1. This space can thus
accommodate two logical qubits, and measuring the plaquette and star operators
allows one to gather information about possible spin- and phase-flip errors
without disturbing the encoded state. A negative plaquette (star) indicates the
presence of one or three $\sigma_x$ ($\sigma_z$) spin errors. The toric code belongs 
to the class of stabilizer codes, and the plaquettes and stars are in that context often referred to 
simply as stabilizer operators.

Notably, the code space $\mathcal{C}$ is the degenerate ground space of the Hamiltonian
\begin{equation}\label{Kitaev Hamiltonian}
 H = -J \sum_s A_s -J\sum_p B_p.
\end{equation}
Here, $J > 0$ is the energy gap, and $A_s$ and $B_p$ are the stars and
plaquettes, respectively, explicitly given by
\begin{eqnarray}
A_s &=& \prod_{i\in\mathrm{adj}(s)}\sigma_x^i, \\
B_p &=& \prod_{i\in\mathrm{adj}(p)}\sigma_z^i,
\end{eqnarray}
where $\mathrm{adj}(s)$ [$\mathrm{adj}(p)$] denotes the set of spins on edges
adjacent to the star (plaquette) $s$ ($p$).
The elementary excitations of the Hamiltonian Eq.~\eqref{Kitaev Hamiltonian} are stabilizer operators with negative eigenvalue and are referred to as `anyons'.

Associated with the two logical qubits encoded in $\mathcal{C}$ are four
string-like operators (products of single-spin $\sigma_x$'s or $\sigma_z$'s)
which wrap around the torus and commute with all plaquettes and stars, but act
non-trivially in the form of logical Pauli $X$ and $Z$ operators on the two
qubits encoded in $\mathcal{C}$. We choose to label the operators such that
$X_1$ is a vertical string on horizontal edges and $Z_1$ is a horizontal string
on horizontal edges. Correspondingly, $X_2$ and $Z_2$ are horizontal and
vertical strings, respectively, on vertical edges (see Fig.~\ref{fig:toric_codes}a).

When the system described by the Hamiltonian Eq.~\eqref{Kitaev Hamiltonian} is
coupled to a noisy environment causing single-spin errors, pairs of anyons are created and can
subsequently move diffusively on the toric surface. Eventually, the creation and
diffusion of anyons leads to a pattern of errors containing undetectable loops
around the torus, acting as unnoticed logical Pauli operators on the code space
$\mathcal{C}$ and therefore corrupting the state contained therein. Measuring the plaquette and star operators to locate anyons reveals some, but
not all information about the underlying error pattern and is generally ambiguous. It is up to an error correction procedure (see Sec.~\ref{sec:simulations and error correction}) to deal
with this problem in a satisfactory way.

The toric code has gained attention due to a series of interesting and advantageous
properties. Namely, the stabilizer operators are local and independent of the system size $L$,
while the code distance grows linearly with $L$. Closely related is the fact that the
ground-state degeneracy is exponentially protected (in $L$) against local perturbations. Quite remarkably, the
toric code is in a sense almost optimal within the class of all CSS codes~\cite{Dennis2002}, even though
the latter contains codes with arbitrarily large stabilizer operators. We will revisit this topic in greater detail
in Sec.~\ref{sec:optimal codes}.

\section{Classical dynamics and numerical simulations}\label{sec:classical dynamics}

\subsection{Classical dynamics from single-spin errors}

Since the Hamiltonian Eq.~\eqref{Kitaev Hamiltonian} does not couple the star
and plaquette operators, we can treat the two corresponding types of anyonic
excitations independently. Furthermore, because the stars are
simply plaquettes on the dual lattice, it is sufficient to study the dynamics of
only one type, e.g., the plaquette anyons. We assume that each spin is coupled
to an auxiliary system that can cause the spin to flip via $\sigma_x$ errors. In
the limit of weak coupling \cite{Davies1974, Alicki2009}, one can derive the
following system of coupled rate equations describing the classical dynamics of
the system \cite{Chesi2010}:
\begin{equation}\label{rate equation}
\frac{d}{dt} p_\mathcal{E}(t) = \sum_{i}\left[\gamma^{in}_{i, \mathcal{E}}
\;p_{x_i(\mathcal{E})}(t) - \gamma^{out}_{i, \mathcal{E}}
\;p_\mathcal{E}(t)\right].
\end{equation}
Here, $p_\mathcal{E}(t)$ is the time-dependent probability to find the system in
the state $|\mathcal{E}\rangle$ obtained by applying $\sigma_x$ errors to all
spins with indices in $\mathcal{E}$, i.e., $|\mathcal{E}\rangle = \prod_{k \in
\mathcal{E}} \sigma_x^k |\psi_0\rangle$, where $|\psi_0\rangle$ is the initial
state of the system. Similarly, $p_{x_i(\mathcal{E})}(t)$ describes
the probability to be in the state $\sigma_x^i |\mathcal{E}\rangle$. Finally,
$\gamma^{in}_{i, \mathcal{E}}$ and $\gamma_{i, \mathcal{E}}^{out}$ are the
transition rates to arrive at or leave the state $|\mathcal{E}\rangle$,
respectively, via a $\sigma_x$-error at the spin with index $i$.

In this work, we will consider two types of error environments. The first one is
a constant error rate model, i.e., we set $\gamma^{in}_{i, \mathcal{E}} =
\gamma^{out}_{i, \mathcal{E}} = const$. In this case, spin-flips are caused
independently of any previously existing anyons and $\sigma_x$ errors. The
second model mimics the coupling to a thermal environment, where the transition
rates are in general energy dependent. Consequently, we set  $\gamma^{in}_{i,
\mathcal{E}} = \gamma(-\omega_{i, \mathcal{E}})$ and $\gamma^{out}_{i,
\mathcal{E}} = \gamma(\omega_{i, \mathcal{E}})$, where $\omega_{i, \mathcal{E}}
= \epsilon_\mathcal{E} - \epsilon_{x_i(\mathcal{E})}$ is the energy difference
between the states $|\mathcal{E}\rangle$ and $\sigma_x^i|\mathcal{E}\rangle$. An
expression for $\gamma(\omega)$ often found in the literature is given by
\begin{equation}\label{gamma_bath_text}
\gamma(\omega)=2 \kappa_n \left|\frac{\omega^n}{1-e^{-\beta\omega}}\right|
e^{-|\omega|/\omega_c}
\end{equation}
and can be derived from a spin-boson model \cite{Leggett1987, DiVincenzo2005}.
Here, $\kappa_n$ is a constant with units $1/{\rm energy}^n$ setting the time
scale, $\beta = 1/k_B T$, with $T$ being the temperature of the bath and $k_B$
denoting Boltzmann's constant, and $\omega_c$ is the cutoff frequency of the
bath. In the following, we set $\omega_c \to \infty$ for simplicity. A bath with $n = 1$ is called `Ohmic', whereas one with
$n \geq 2$ is called `super-Ohmic'. Only the former case is considered in this work. Unless otherwise stated, all energies will be expressed in
units of $k_B T$. Consequently, the unit of time is $(\kappa_1 k_B T)^{-1}$.

\subsection{Simulations and error correction}\label{sec:simulations and error correction}

Clearly, it is impossible to solve Eq.~\eqref{rate equation} analytically for
meaningful system sizes, because the number of states $p_\mathcal{E}$ grows
exponentially with $L^2$. We thus have to stochastically simulate the system and
obtain the quantities of interest, such as the number of anyons or the
expectation values of the logical operators, by averaging over many (typically
several thousand) instances. 
In greater detail, the iteration of the simulation at time $t$ consists of these
steps: (i) Calculate all unnormalized single spin-flip probabilities $p_i =
\gamma(\epsilon_\mathcal{E} - \epsilon_{x_i(\mathcal{E})})$, then obtain from
them the total spin flip rate $R = \sum_i p_i$. (ii) Draw the time $\Delta t$
until the next spin flip from an exponential distribution with rate $R$. (iii)
Calculate and record all quantities of interest for time sampling points lying
in the interval $[t, t+\Delta t]$. Namely, these quantities are the number of
anyons, the number of $\sigma_x$ errors, and the uncorrected and error-corrected
(see below) logical operators $Z_1$ and $Z_2$. (iv) Determine a random spin
according to the probabilities $p_i/R$, flip it, and set $t$ to $t + \Delta t$.

The error correction step in the toric code consists of pairing up all detected
anyons and then annihilating them by connecting each pair with a string of
errors
from one partner to the other. The pairing is usually chosen such that all
anyons are annihilated with the smallest total number of single-spin operators.
This is known as the minimal-weight perfect matching and can be found  in
polynomial time with the help of the `blossom' algorithm due to Edmonds \cite{Edmonds1965}.
The runtime complexity of this algorithm has been improved several times since
its discovery. We are employing the library \texttt{Blossom V} \cite{Kolmogorov2009} which implements
the latest version running in $O(mn\log n)$ time, where $n$ is the number of anyons
(vertices) and $m$ the number of connections (edges) between them. 

In order to find the true matching with minimal weight, one in principal would
need to choose the set of edges to include all connections from every anyon to
every other. However, since the size of this set grows quadratically with the
number of anyons $n$, the overall scaling of the matching algorithm becomes
$O(n^3 \log n)$ which is infeasible for large $n$. We therefore first perform a
Delaunay triangulation in negligible $O(n \log n)$ time using the library
\texttt{Triangle} \cite{Shewchuk1996}. The result is that only anyons close to each other are
connected using a number of edges linear in the number of anyons. It turns out
that this is an excellent approximation yielding results that are nearly
indistinguishable from those obtained from a matching over the complete graph.

Within the paradigm of active error correction, where the anyons are detected
and corrected periodically on sufficiently small time intervals, the encoded
state can be kept free of logical errors almost indefinitely. However, since we
are interested in the use of the toric code as a passive quantum memory, we are
mostly concerned with the lifetime $\tau$ of the encoded information in a
scenario where error correction is only performed once at readout. Hence,
whenever we show plots of the `error-corrected' logical operators decaying as a
function of time, we thereby refer to their value if error correction had been
performed at that time, without actually performing it. We then define the
lifetime of the system as the time it takes for the expectation values of the
error-corrected logical operators to decay to $90\%$ of their initial value.

\section{Random lattices}\label{sec:random lattices}

In this section, we study the error thresholds of a family of
models obtained by randomly modifying the toric code in a way that preserves its
basic features. We first describe how we create our random lattices and then
present and discuss the results of the simulations within the context of optimal
quantum codes.

\begin{figure}
\centering
\includegraphics[width=0.8\columnwidth]{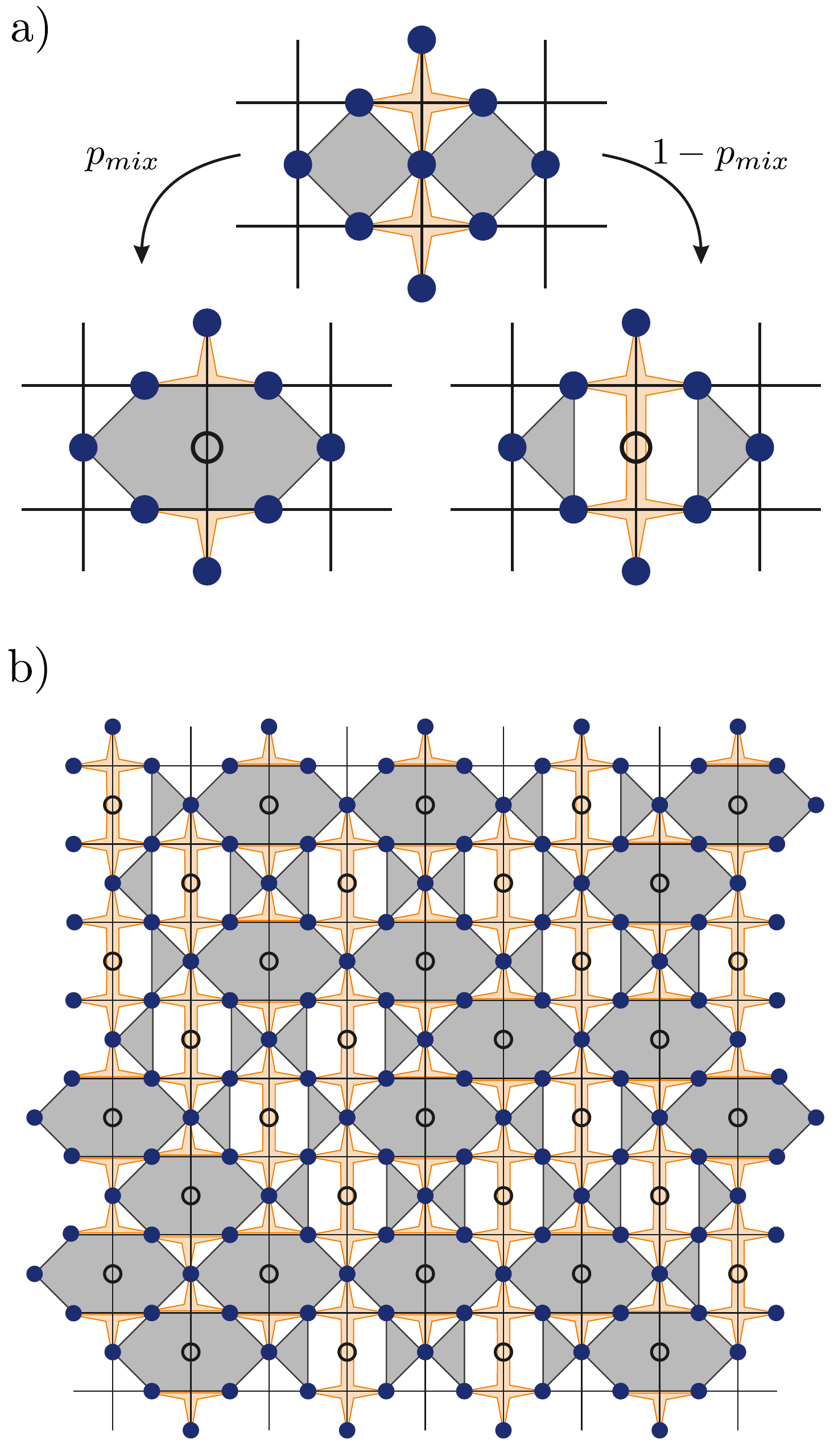}
\caption{(color online) Modifying the stabilizer operators. a) When removing a spin (empty circle), we choose between two ways of adapting the affected stabilizer operators. With probability $p_{mix}$ we join the two plaquette operators to one large 6-body operator and reduce the two stars from 4- to 3-body operators. Alternatively, with probability $1-p_{mix}$, we perform the dual operation, namely we define two 3-body plaquettes and one large 6-body star. Spins and operators not affected by removing the central spin are not shown in this example. b) Typical $8\times 8$ subregion of a (larger) random lattice with $p_{mix} = 0.5$. Logical operators as well as some spins and operators at the edges with the rest of the lattice are not shown.}\label{fig:random_modification}
\end{figure}

\subsection{Generating the lattices}

Starting from the toric code on an $L \times L$ lattice, we remove $\frac{1}{2}
L^2$ spins at specific and regularly distributed `defect' locations. The
structure of the defect pattern can be easily understood from Fig~\ref{fig:toric_codes}b. Basically,
every second vertical edge is labelled as a defect, with the first defect
of each row alternatingly being created on the first or second vertical edge of that row.
Note that the height and width of the grid both must be even in oder for this
procedure to be consistent with the periodic boundary conditions. We now have to
modify all plaquettes and stars as well as the logical Pauli $X$ and $Z$ operators
such that all original commutation relations remain
unaltered.

Let us start with the logical operators. Clearly, both $X_1$ and $Z_1$
(with single-spin operators exclusively on horizontal edges) are unaffected by the
introduction of defects on vertical edges of the lattice. However, the pair of
operators in the original code acting on the second encoded qubit is defined on
vertical edges and thus needs to be adapted. Clearly, the operators must remain
connected strings wrapping around both dimensions of the torus. The `zig-zag'
pattern shown in Fig.~\ref{fig:toric_codes}b achieves this with the smallest increase in the number
of single-spin operators. It is straightforward to verify that these new $X_2$
and $Z_2$ operators, together with the unaltered pair acting on the first
encoded qubit, indeed fulfill all original commutation relations between each
other. 

We now discuss the modification of the plaquette and star operators. Removing
one spin, i.e., creating one defect, affects exactly two adjacent plaquettes and
two adjacent stars. We will consider two possible ways of dealing with this
situation (cf. Fig.~\ref{fig:random_modification}). We can either (i) define two restricted 3-body
plaquette operators and one `vertical star' operator consisting of the product of the remaining 6
single-spin $\sigma_x$ operators, or (ii), perform the dual operation, namely
defining one large 6-body `horizontal plaquette' and two restricted 3-body
stars. The two ways of modifying the original operators are depicted in Fig~\ref{fig:random_modification}a.
It is relatively easy to see that these new 3-body and 6-body operators remain
mutually commuting, and furthermore commute with all modified logical Pauli
operators just as in the original code. Note that also the dimension of the code
space is left unchanged since it generally only depends on the genus of the
surface covered by the anyon operators \cite{Kitaev2003}. We can now create a random
lattice by choosing at each defect site to create a 6-body plaquette with
probability $p_{mix}$ and two 3-body plaquettes with probability $1 - p_{mix}$.
See Fig. \ref{fig:random_modification}b for a typical example. The special case $p_{mix} = 0$ corresponds to
a regular lattice of 3-body plaquettes and 6-body stars, whereas $p_{mix} = 1$
conversely yields a regular lattice of 6-body plaquettes and 3-body stars.
In both cases, the 3-body operators are the vertices of an underlying hexagonal lattice, whereas the 
6-body operators form the vertices of its dual, the triangular lattice.

\subsection{Results}

The randomization of the lattice geometry described above creates a difficulty in employing the Delauny triangulation, because the graph is now irregular. 
To overcome this problem we have replaced this step by a
breadth-first search performed on each anyon. This procedure connects every
anyon to at most $k$ of its nearest neighbors, where distance is measured not in
a Euclidean sense, but as the number of errors in a connecting string. For
constant $k$, this requires a runtime of $O(n)$, where $n$ is the number of
anyons. We have found that $k = 10$ is an excellent approximation to $k =
\infty$ and have used this value in all calculations.

We have performed a series of Monte Carlo simulations to determine the
critical fraction of errors $f_{cr}^Z$ independent of any form of anyon
dynamics (see Appendix \ref{app:random lattices ohmic} for additional results in the case of thermal errors). If the probability for each spin to independently be
affected by a $\sigma_x$ error becomes larger than this critical value in the
limit $L\rightarrow\infty$, the error correction scheme undergoes a
transition from performing fully accurate error recovery to randomly guessing
the error-corrected state with the lowest possible success rate of $50\%$. We
can determine $f_{cr}^Z$ by plotting the expectation values of the
error-corrected logical $Z$ operators as a function of the error probability $f$
for different lattice sizes and observe at which value of $f$ the curves
intersect~\cite{Note1}.

\begin{figure}
\centering
\includegraphics[width=0.95\columnwidth]{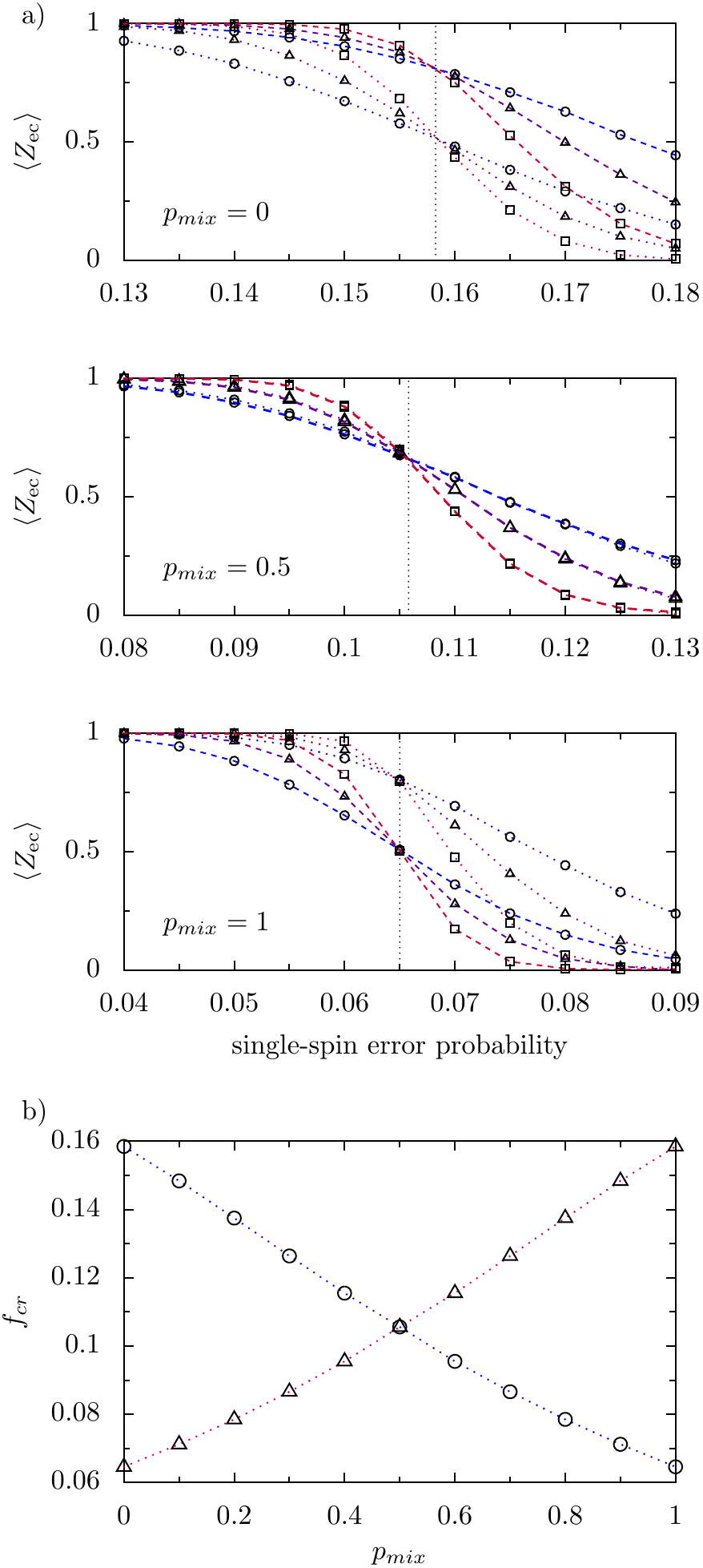}
\caption{(color online) Critical error thresholds of random models. a) Three example plots of data used to determine the critical error thresholds. Each plot shows, for a specific value of $p_{mix}$, the expectation values of the logical $Z_1$ (dotted lines) and $Z_2$ (dashed lines) operators for grid sizes $L=32\;({\rm circles}),\;64\;({\rm triangles}),\;{\rm and}\; 128\;({\rm squares})$. The vertical dotted lines indicate the position of the error threshold. Data points are obtained by bootstrapping 1000 sample values, each of which is obtained by averaging over 200 random error distributions on a single instance of a random lattice. b) Error thresholds [as determined in a)] of $Z$ (circles) and $X$ (triangles) operators as a function of $p_{mix}$. The dotted lines are guides to the eye.}\label{fig:metr threshold}
\end{figure}

Fig.~\ref{fig:metr threshold}a displays typical results for a few different values of $p_{mix}$. We observe
that the expectation values of $Z_1$ and $Z_2$ in general have a different
dependence on the error probability $f$. This is due to the fact that the fraction of 6-body plaquettes determines the average number of spins required to form
a loop around the horizontal direction of the torus, whereas this number is constant for the vertical direction.
Not surprisingly, $Z_1$ and $Z_2$ decay identically for a 50 per cent mixing of 3- and 6-body plaquettes.
Note that, despite the typically unequal decay of $Z_1$ and $Z_2$ as a function of $f$, the error thresholds for the two operators are always identical for a given value of $p_{mix}$. This is consistent with the general
understanding that the correctability of the memory as a whole is related to the phase of a corresponding random-bond Ising model \cite{Dennis2002}. Indeed, our numerically determined thresholds for the regular 3-body and 6-body
plaquette lattices agree well with the recently calculated multicritical points in spin glass models on hexagonal and triangular lattices, respectively \cite{Ohzeki2009}. Specifically, we find $f_c \approx 0.1585$ for $p_{mix} = 0$ (theoretical value: $f_c = 0.1640$) and  $f_c \approx 0.0645$ for $p_{mix} = 1$ (theoretical value: $f_c = 0.0674$). The discrepancy of about $4-5\%$ between the numerical and theoretical thresholds is of the same size as in the case of the toric code on a square lattice (where we had found $f_c \approx 0.1055$ as compared to the theoretical value $f_c = 0.1092$) \cite{Chesi2010} and can generically be attributed to the failure of the minimal-weight perfect matching close to the threshold. Interestingly, our numerical results suggest that the thresholds of the toric code and our random lattice models with $p_{mix} = 0.5$ are identical.

We show in Fig.~\ref{fig:metr threshold}b the critical fraction of errors $f_{cr}^Z$ for the
logical $Z$ operators determined in the way described above as a function of the
lattice mixing probability $p_{mix}$. Since the plaquette and star lattices are
dual to each other (to every 6-body plaquette correspond two 3-body stars and
vice versa), the critical fraction $f_{cr}^X$ of $\sigma_z$ errors for which
error correction of the logical $X$ operators fails (also plotted in Fig.~\ref{fig:metr threshold}b) is simply given by
\begin{equation}\label{eq:fcrx}
f_{cr}^X(p_{mix}) = f_{cr}^Z(1 - p_{mix}). 
\end{equation}
At equal mixing, i.e., $p_{mix} = 0.5$, the threshold values are given by $f_{cr}^X(0.5) = f_{cr}^Z(0.5) \approx 0.1055$.

Consequently, one of the thresholds for the two different types of Pauli operators, either $f_{cr}^X$ or $f_{cr}^Z$ is always smaller than or equal to the threshold of the toric code. Our random lattices thus bear no advantage
over the latter in the case of a uniform error model, where $\sigma_x$ and $\sigma_z$ errors occur with the same probability. The situation is different, however, for biased noise. If bit-flips and phase-flips are created with different probabilities, we can make use of the asymmetry in the error thresholds for $p_{mix} \neq 0.5$. Assuming, for instance, that $\sigma_x$ errors are more frequent than $\sigma_z$ errors would lead to an overall lifetime decrease of encoded states in the toric code due to the shorter lifetimes of the logical $Z$ operators. However, starting from a random model at $p_{mix} = 0.5$, a decrease in $p_{mix}$ will lead to an increase of the $Z$ lifetimes and a decrease of the $X$ lifetimes. If the error frequencies are not too different, the lifetimes will become identical at some value $0 \leq p_{mix} < 0.5$ and will be larger than the overall lifetime of the toric code. We thus conclude that the random lattices can be employed to increase the lifetime of encoded states compared with the toric code on a square lattice in the presence of biased noise. While these lattices require both error probabilities to fall in the range $0.0674 \leq p \leq 0.1640$ (and below the boundary in Fig.~\ref{fig:optimal bound}, see next section), it should in principle be possible to extend this range to $0 \leq p \leq 0.5$ by defining stabilizers with more than 6 single-spin operators in a similar fashion.

\subsection{Relation to optimal quantum codes}\label{sec:optimal codes}

We now discuss our random lattices within the context of optimal quantum coding. It is well known that, assuming a 
biased constant error model, there is an upper bound on the fraction of logical qubits $k$ and physical qubits $n$ that encode them, valid for all
CSS codes. This bound is given by \cite{Steane1996a, Dennis2002, Gottesman2003}
\begin{equation}\label{eq:css bound}
k/n \leq 1 - H(p_x) - H(p_z),
\end{equation}
where $H(x) = -x \log_2 x - (1 -x) \log_2(1 - x)$ is the Shannon entropy, and $p_x$ and $p_z$ are the probabilities for a single spin to be
affected by a $\sigma_x$ and $\sigma_z$ error, respectively. The bound Eq.~\eqref{eq:css bound} can be motivated with the following intuitive argument.

An ideal CSS code would be able to detect for each physical qubit if it was suffering from a $\sigma_x$ or a $\sigma_z$ error. Assuming that these errors are uncorrelated,
the number of classical bits required to store this information is asymptotically given by $n H(p_x) + n H(p_z)$. If we are to store the same information in 
qubits instead of bits, the Holevo bound \cite{Nielsen2000a} requires the usage of at least as many qubits to do so. Since our optimal CSS code needs to store the information of $k$ encoded qubits, as
well as all possible occurrences of errors, we have
\begin{equation}
n \geq k + n H(p_x) + nH(p_z).
\end{equation}
Dividing by $n$ and rearranging the terms yields the desired bound Eq.~\eqref{eq:css bound}.

\begin{figure}
\centering
\includegraphics[width=0.95\columnwidth]{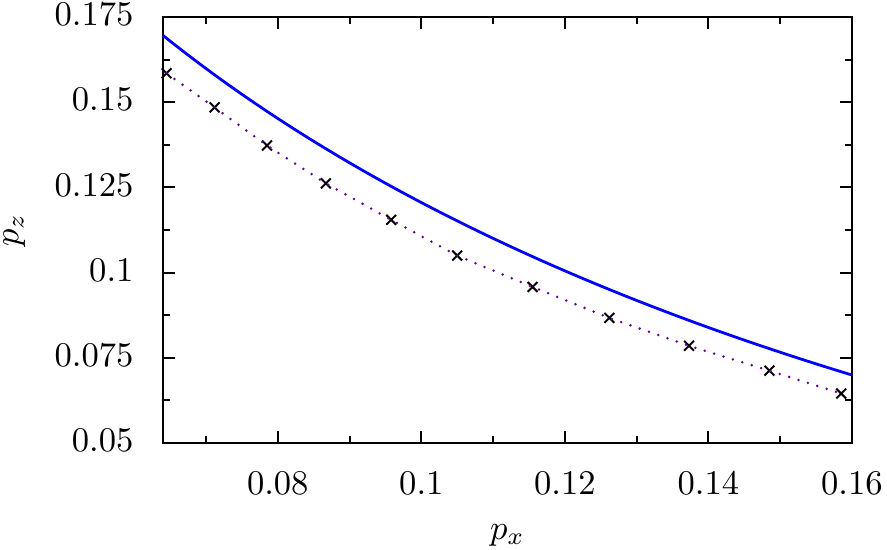}
\caption{(color online) Theoretical upper bound on biased noise correctable by CSS codes. Given an error model of independent $\sigma_x$ and $\sigma_z$ errors occurring with constant probabilities $p_x$ and $p_z$, respectively, there exists no CSS code able to cope with pairs of error probabilities lying above the solid line given by the zero-contour of Eq.~\eqref{eq:css bound}. The crosses are the numerically determined pairs of thresholds of the random models for $p_{mix} = 1$ down to $p_{mix} = 0$ from left to right. The dotted line is a guide to the eye.}\label{fig:optimal bound}
\end{figure}

For unbiased errors with $p = p_x = p_z$, the right-hand side of Eq.~\eqref{eq:css bound} becomes zero at $p \approx 0.110028$, implying that there cannot be any CSS code coping with an error rate larger than this value. Quite remarkably, the critical error probability for the toric code has been determined to be $f_{cr} = 0.109187$ \cite{Ohzeki2009}. This is astonishingly close to the upper bound, especially when
taking into account that all stabilizers are local 4-body operators. Moreover, inserting the error thresholds for the regular 6-body plaquette lattice and its dual lattice of 3-body stars ($p_{mix} = 1$) into the right-hand side of Eq.~\eqref{eq:css bound} evaluates to
\begin{equation}
 1 - H(0.0674) - H(0.1640) \approx 6\times 10^{-5},
\end{equation}
which is virtually zero, indicating that the code is close to optimal for this particular biased error model. Due to the symmetry of Eq.~\eqref{eq:css bound} with respect to the error probabilities and the duality of the triangular and hexagonal lattices, the same argument holds for a lattice with 3-body plaquettes and 6-body stars ($p_{mix} = 0$) with the values of $p_x$ and $p_z$ exchanged. With our random lattices, we can thus continuously interpolate between two optimal models by changing $p_{mix}$. This suggests that the random models are optimal for all values of $p_{mix}$, in the sense that for every $0.0674 \leq p_x \leq 0.1640$ there is a random model with a theoretically (close to) maximal possible threshold for $p_z$. The results plotted in Fig.~\ref{fig:optimal bound} strongly support this claim. The solid line is the zero-contour of the upper bound Eq.~\eqref{eq:css bound} and the crosses are the threshold pairs of the random lattices determined numerically. Note that the numerical data is within the typical $5\%$ distance of theoretical bound. This can once again be explained by the failure of the minimal-weight error correction algorithm close to the thresholds. This observation, together with the knowledge from theory that the models are virtually optimal for $p_{mix} = 0$ and $p_{mix} = 1$ leads us to conjecture that the random models are virtually optimal for all values of $p_{mix}$. However, carrying out a theoretical study in the fashion of Ref.~\cite{Ohzeki2009} is outside of the scope of the present work and is deferred for future research.

\section{Random onsite potentials}\label{sec:random potentials}

This section is devoted to the study of the classical dynamics of anyons in the regular toric code on a square lattice, but with randomly modified anyon onsite energies. We are also particularly interested in the case where long-range anyon-anyon interaction is present, as this has been shown to generally enhance the lifetime of the memory due to the suppression of the anyon density with increasing system size \cite{Chesi2010}. For this, it is convenient to introduce the new stabilizer operators $n_s = (1 - A_s)/2$ and $n_p = (1 - B_p)/2$, where $A_s$ and $B_p$ are the usual star and plaquette operators, respectively. These operators are zero in the absence of an anyon on the respective site, and equal to 1 otherwise. The more general Hamiltonian can then be written as
\begin{equation}\label{H_interacting}
H = \frac12 \sum_{p p'} U_{pp'} n_p n_{p'} + \frac12 \sum_{s s'} V_{ss'} n_s n_{s'},
\end{equation}
where $U_{pp'}$ and $V_{ss'}$ contain the onsite energy and repulsive anyon interaction terms. Since in this model plaquette and star anyons are still independent, we can set $V_{ss'} = 0$ and note again that all results for the plaquette anyons hold equally for the stars. We then set \cite{Chesi2010}
\begin{equation}\label{Upp_long_range}
U_{pp'}=2J_p \delta_{pp'}+\frac{A}{(r_{pp'})^\alpha} (1-\delta_{pp'}),
\end{equation}
where $J_p$ is the onsite energy of an anyon on the plaquette with index $p$, $A$ is the interaction strength, $r_{pp'}$ is the shortest distance on the torus between plaquettes $p$ and $p'$, and $0 \leq \alpha < 2$ is the (long-range) interaction exponent. The onsite energies $J_p$ are chosen randomly from a distribution with mean zero in order to discriminate effects caused by the randomness from effects potentially caused by the system having a non-zero mean gap.

\begin{figure}[!t]
\centering
\includegraphics[width=0.95\columnwidth]{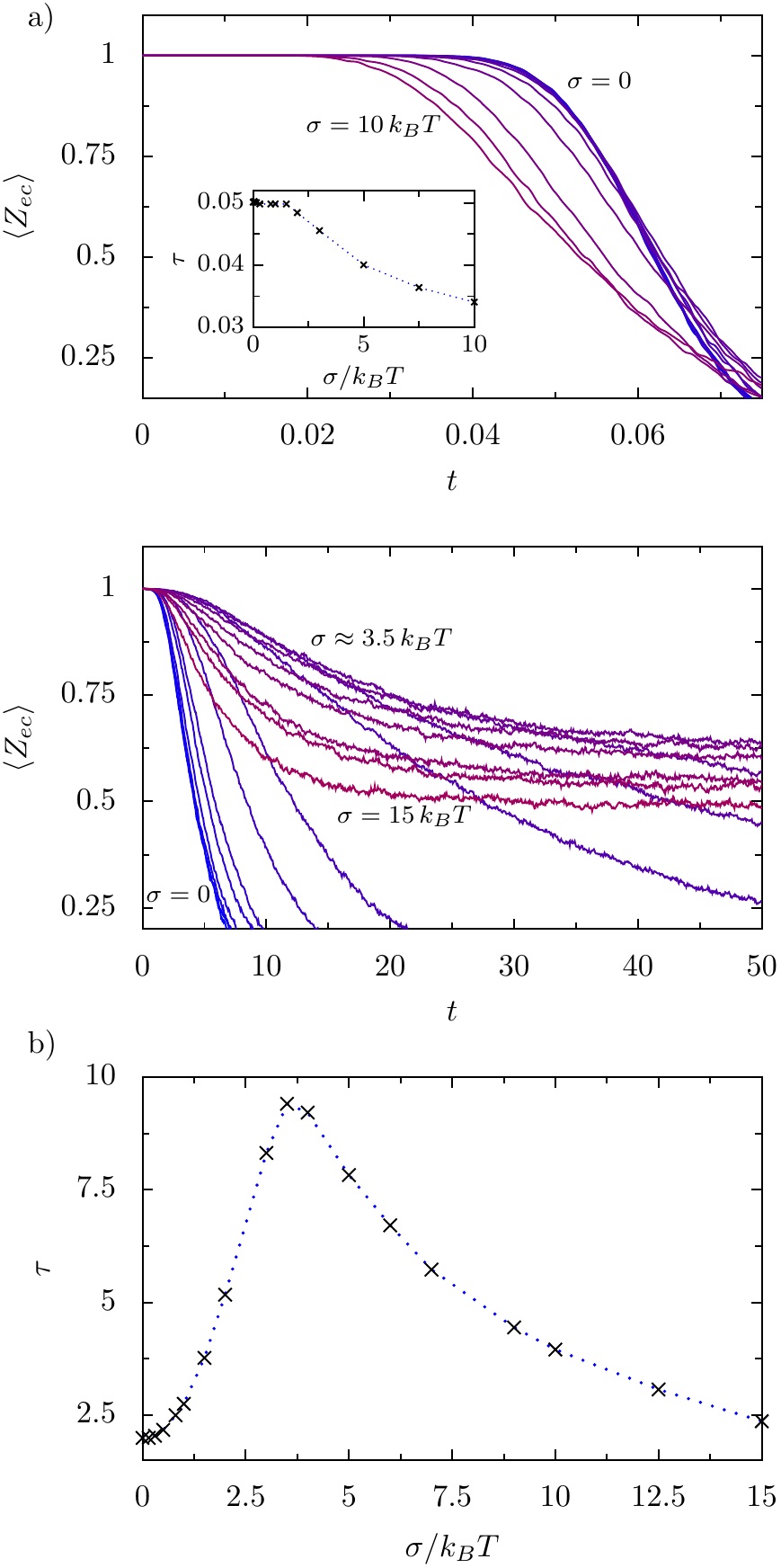}
\caption{(color online) Influence of disorder on the memory lifetime. a) Time evolution of error corrected $Z$ operators for a non-interacting (top) and interacting (bottom, $\alpha = 0$, $A = 0.5$) model. Both systems are of size $32 \times 32$ unit cells ($2\times 32^2$ spins) and are coupled to an Ohmic bath at temperature $T = 1$. The different curves display $\langle Z_{ec}(t)\rangle$ for different disorder strengths $\sigma$ of Ising-like randomness with onsite energies $J_p = \pm \sigma$. In the non-interacting system, $\sigma$ is increased from $0$ to $10\, k_B T$ as indicated in the panel. The inset displays the lifetime of the memory, i.e., the time at which $\langle Z_{ec}\rangle$ hits $0.9$, as a function of $\sigma$. The disorder strengths examined in the interacting case have been chosen as $0 \leq \sigma \leq 15\, k_B T$ (see main text and labels in the panel). b) The lifetimes of the interacting model extracted from the curves of the lower panel in a). The dotted line is a guide to the eye. \vspace{0.1cm}}\label{fig:compare_t90_nonint_int}
\end{figure}

We focus on the case of constant interaction, i.e., $\alpha = 0$, and Ising-like randomness, meaning that the $J_p$ are chosen from $\{-\sigma, +\sigma\}$ with equal probabilities. We refer to $\sigma \geq 0$ as the disorder strength. This model is interesting mostly for two reasons. First, it is the most convenient system incorporating randomness with respect to numerical simulation. Since the interaction is constant and thus simply depends on the total number of anyons, only six different single-spin flip rates need to be updated at each iteration step, depending on the number and configuration of adjacent anyons and onsite energies, respectively. Second, this simple model already displays all dynamical effects also present in more complicated systems (e.g., $\alpha \neq 0$ and Gaussian distribution of $J_p$'s, see Appendix~\ref{app:random gauss}) and thus serves as an ideal playground for studying these effects. Naively, one would expect that the presence of negative onsite energies in the system simply favors the creation of anyons and is thus always disadvantageous for the lifetime of the memory. While this is indeed true for a non-interacting system, we find a regime in the interacting case where, quite surprisingly, the lifetime is enhanced for increasing disorder strength.

Fig.~\ref{fig:compare_t90_nonint_int} presents the results for the two cases. The non-interacting system is stable against disorder strengths that are roughly equal to the temperature but then decays for larger $\sigma$. This can be understood easily from the detailed balance condition satisfied by the rates Eq.~\eqref{gamma_bath_text}: The ratio of the creation and annihilation rates of a pair of anyons on two sites with negative onsite energy is given by $\gamma(2\sigma)/\gamma(-2\sigma) = \exp(2\beta\sigma)$, which becomes large for $\sigma$'s exceeding the temperature. It is thus exponentially more likely for a pair of anyons to be created than annihilated for $\sigma > k_B T$, thereby quickly cluttering the system with anyons and crossing the critical fraction of errors. This situation changes completely in the presence of interactions between the anyons. The data shown in Fig.~\ref{fig:compare_t90_nonint_int} displays a steep increase in the lifetime as a function of the disorder strength, peaking at around $\sigma \approx 3.5\,k_B T$ for that particular system, followed by a slower decay.

\begin{figure}[!tb]
\centering
\includegraphics[width=0.95\columnwidth]{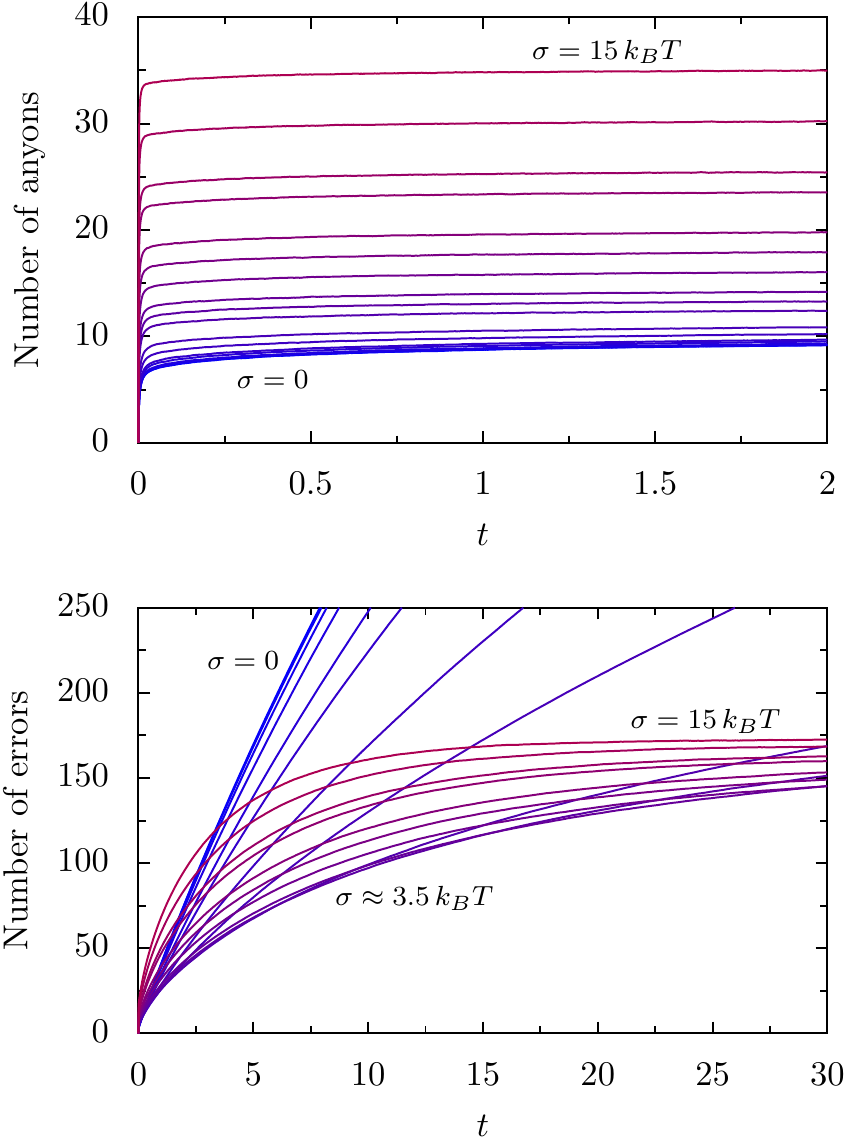}
\caption{(color online) Number of anyons (top) and number of single-spin errors (bottom) as a function of time in an interacting system ($\alpha = 0, A = 0.5\, k_B T$) of size $32\times 32$ coupled to an Ohmic bath. The strength of the Ising-like randomness is increased from $\sigma = 0$ to $\sigma = 15\, k_B T$ as indicated in the panels (see also main text).}\label{fig:interacting_nf}
\end{figure}

We can shed some light on this effect by additionally looking at the number of anyons and the number of errors, as shown in Fig.~\ref{fig:interacting_nf}. For any fixed value of $\sigma$, the number of anyons again increases quickly but then saturates almost instantaneously at the equilibrium value. At this point, creating a new pair of anyons costs an energy penalty due to the repulsive interaction that can no longer be compensated by the negative onsite potentials. One can clearly see that the equilibrium number of anyons increases linearly with $\sigma$, which implies that the corresponding enhanced memory lifetimes cannot be explained by a suppression of the anyon density. 

However, the error creation rate (i.e., the slope of the curves in the lower panel of Fig.~\ref{fig:interacting_nf})  exhibits a pronounced minimum at the same value $\sigma\approx 3.5\,k_B T$ that also yields the maximal lifetime. Such an initial decrease in the error rate despite an increasing number of anyons can only be consistently explained by a suppression of the anyon diffusion. 
For disorder strengths $\sigma \gg k_B T$, processes that create anyons on two positive sites or that move an anyon from a negative to a positive site are exponentially suppressed. The positive sites thus effectively act as infinite barriers that greatly reduce the mobility of the anyons, and the encoded state is solely destroyed by diffusing anyons restricted to the negative sites~\cite{Note2}. As the disorder strength is lowered, two competing effects come into play. On the one hand, the number of anyons is reduced linearly. On its own, this would lead to a longer lifetime due to the presence of fewer diffusing anyons. On the other hand, the barriers separating the regions of negative sites are lowered, which facilitates the diffusion across longer distances and promotes a reduction in lifetime. The observed maximum in the memory lifetime can thus be understood as a tradeoff between having few but relatively freely moving anyons for $\sigma \ll k_B T$, and more but very restricted anyons for $\sigma \gg k_B T$. The interaction merely plays the role of restricting the anyons to a small enough (for $\sigma \lesssim k_B T$) and constant number. Appendix~\ref{app:supporting} contains results that further support the picture described above.

\section{Quantum dynamics}\label{sec:quantum dynamics}

\subsection{Error model}

Consider the toric code Hamiltonian, perturbed by a magnetic field of strength $h$. For concreteness, let us choose this to be of the form,
\be
H = \sum_p J_p n_p + \sum_s J_s n_s + h \sum_i \sigma_x^i.
\ee
The effects of such a perturbation have been studied using the methods of Refs. \cite{Wootton2011, Stark2011}, where it was noted that, since the $\sigma_x^i$ do not commute with the $n_p$, the perturbation will have the effect of creating, annihilating and transporting plaquette anyons. For $h\ll J_p$ all these effects apart from the transport are suppressed by the energy gap, allowing the system to be modelled as the following many-particle quantum walk Hamiltonian,
\bq \nonumber
H_p &=& \sum_{p,p'} M_{p,p'} t_{p,p'} + U \sum_p n_p (n_p-1), \\\label{M}
M_{p,p'} &=& \delta_{\langle p,p' \rangle} h + \delta_{p,p'} J_p.
\eq
Here $\delta_{\langle p,p' \rangle}=1$ only when the plaquettes $p$ and $p'$ share a spin. The operator $t_{p,p'}$ maps a state with an anyon on the plaquette $p$ to that with the anyon moved to $p'$, and annihilates any state without an anyon initially on $p$. Since the anyons are hardcore bosons, we are interested in the case of $U\rightarrow \infty$.

This effective description in terms of quantum walks of anyons holds also for a more general magnetic field and other local perturbations. The effects of anyonic braiding occur at a higher order of perturbation theory than those of this effective description, and hence may be ignored. The dynamics of the plaquette and vertex anyons can therefore be considered separately. Since they are dual to each other, once again only the plaquette anyons are considered here without loss of generality.

The Hamiltonian $H_p$ is difficult to solve in general. However, note that the dynamics of $H_p$ are driven by the matrix $M$, i.e., the Hamiltonian for a single particle walk. Hence, by considering the case of a single anyon, important aspects of the behavior for the many-particle walks can be determined. It is this approach that is taken here. The Hamiltonian $M$ is applied to a single anyon, initially placed on an arbitrary plaquette of the code. The motion of the anyon can be characterized by the time evolution of its standard deviation, $\Delta$. Since finite values of the system size $L$ must be used in the numerics, the walks will, at some point, interact with the boundary. In order for this effect to be ignored, the run time of the walks is limited to ensure this interference always remains negligible.

\begin{figure}[t]
\begin{center}
{\includegraphics[width=8cm]{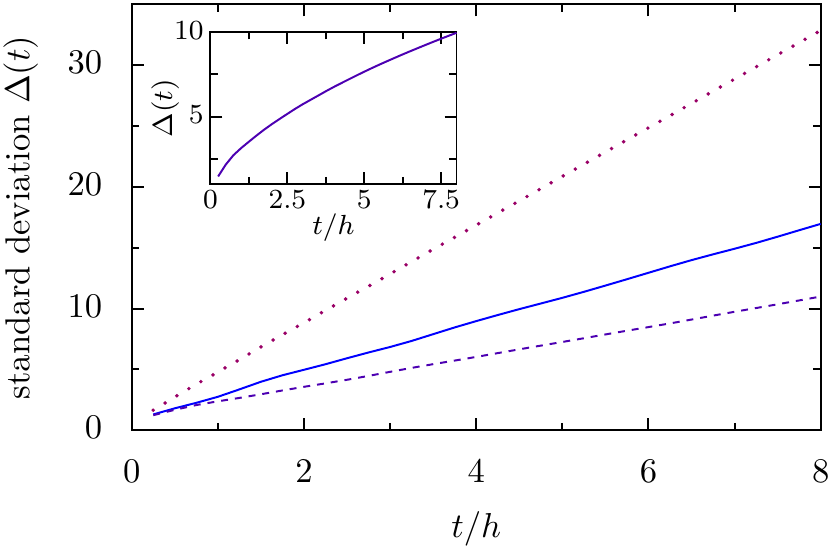}}
\caption{\label{fig:quantum_uniform_J} Time evolution of the standard deviation $\Delta(t)$ of a single quantum walker on lattices with uniform couplings $J_p = J$ for all $p$. The different curves correspond to a square lattice (solid), 3-body plaquette lattice (dashed), and 6-body plaquette lattice (dotted). Inset: $\Delta(t)$ on the random lattices from Sec.~\ref{sec:random lattices} with $p_{mix} = 0.5$, where each point has been averaged over 1000 samples.}
\end{center}
\end{figure}

The behavior of $\Delta$ over time for walks on square, triangular and hexagonal lattices for which all $J_p$ are uniform can be found in Fig. \ref{fig:quantum_uniform_J}. In each case the standard deviation of the distance increases linearly with time, demonstrating the ballistic motion expected from quantum walks when no disorder is present.

The ballistic motion caused by the field is highly damaging to the quantum information stored within the code. Suppose that the toric code initially has some density $\rho$ of anyon pairs, due perhaps to noisy preparation of the state or interaction with the environment. If $\rho$ is sufficiently small then the pairs will be far apart, allowing error correction to be performed reliably. However, when the perturbation is present this will only remain true for a finite lifetime $\tau$, after which the motion of the anyons prevents them from being paired reliably. This occurs when they have moved a distance comparable to the average distance between pairs, and hence when $\Delta(\tau) \sim 1/\sqrt{\rho}$. Since $\Delta(t)$ grows linearly with time for ordered quantum walks (cf. Fig.~\ref{fig:quantum_uniform_J}), the quantum memory will fail within a time of order $\tau = O(1/\sqrt{\rho})$. Mechanisms which slow down the anyons are therefore favorable to the quantum memory, since they lead to longer lifetimes. It is this effect that is expected to emerge from the disorder.

\subsection{Random lattices}

Let us now introduce disorder by using the random lattice of Sec.~\ref{sec:random lattices} while still keeping the $J_p$ (and $J_s$) uniform, all taking the same value $J$. Specifically the case of $p_{mix} = 0.5$ is considered, to maintain the symmetry between plaquette and star anyons. The behavior of the standard deviation of the distance moved by a single walker is shown for this lattice in the inset of Fig. \ref{fig:quantum_uniform_J}. Rather than increasing linearly with time $t$, as in the ordered case, it is found that $\Delta(t)$ grows with the square root of $t$. The motion of a quantum walker is therefore diffusive rather than ballistic in this case. As such, the random lattice leads to a significant slowing of the anyon motion, increasing the lifetime to $\tau = O(\rho^{-1})$ (note that we always have $\rho \leq 1$). It is possible that the random lattice also induces Anderson localization \cite{Anderson1958}, in which case the lifetime will be increased further, but the system sizes which may be probed are too small for this to be evident.

\subsection{Random lattices together with $J$ disorder}

It is known that, when disorder in the $J_p$ couplings is present in the toric code, Anderson localization is induced \cite{Wootton2011, Stark2011}. This effect exponentially suppresses the motion of the walkers, and causes the standard deviation of the distance to converge to a constant value. We therefore have $\tau \rightarrow \infty$, i.e., the memory stays stable against the perturbation for an arbitrarily long time.  It is now important to determine whether the combination of randomness in both the lattice and the $J_p$ couplings enhances or diminishes this effect.

To study this, disorder in the $J_p$ couplings are considered. Specifically, each $J_p$ randomly takes either the value $J-\sigma$ or $J+\sigma$ with equal probabilities. The value $J$ is unimportant, but the ratio of $\sigma/h$ characterizes the strength of the disorder in comparison to the magnetic field. Guided by the numerical results of \cite{Wootton2011}, we consider here disorder of strength $\sigma/h = 250$ to ensure that the localization effect is observed for moderately sized systems.

\begin{figure}[t]
\begin{center}
{\includegraphics[width=8cm]{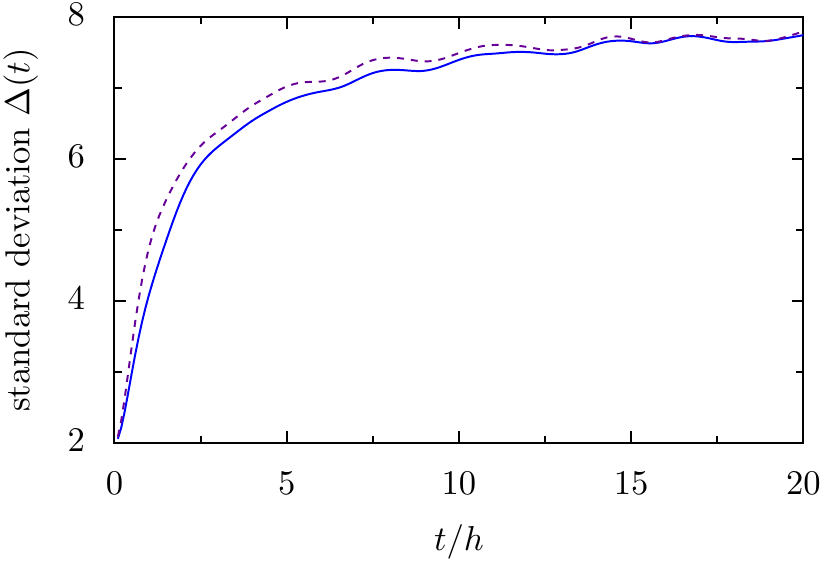}}
\caption{\label{randomlatticeandJ} Time evolution of the standard deviation $\Delta(t)$ for a single anyonic walker with disorder in the $J_p$ of $\sigma/h = 250$, on both a square lattice (solid line) and a random lattice (dashed line) with $p_{mix}=0.5$. Each point has been averaged over 1000 samples.}
\end{center}
\end{figure}

In Fig. \ref{randomlatticeandJ}, the time evolution of the standard deviation is shown for the case of $J$ disorder on a square and a random lattice. In both cases the localization effect is seen, with the walk unable to move far beyond a few times the length scale separating neighboring vertices. The walks with and without the lattice disorder give very similar results, especially at longer times. The effect of localization in the random lattice therefore seems the same as that of the square, without significantly enhancing or diminishing the effect.

\section{Summary}\label{sec:conclusions}

We have studied the influence of quenched disorder on the incoherent (classical) motion of anyons in modified forms of the toric code. We have first described a class of random models that can be obtained from the toric code by  removing a regular sublattice of spins, and then for each defect site randomly choosing one of two ways to adapt the affected stabilizers with a probability $p_{mix}$. The critical fractions of independent errors at which these codes become uncorrectable have then been determined numerically as a function of $p_{mix}$. We have shown that in the presence of biased noise, where bit flips and phase flips occur at different probabilities, the models based on random lattices can tolerate higher thresholds than the toric code in one type of errors, given the other type is correspondingly lower. These thresholds have been demonstrated to be close to the upper bound correctable by any CSS code. Second, we have studied the toric code subject to randomness in the onsite potentials. Specifically, we have demonstrated that in the presence of repulsive long-range interaction between anyons, there is a pronounced maximum in the lifetime of encoded states as a function of disorder strength. This effect has been attributed to a reduction of anyon diffusion due to the sites with positive onsite energy acting as barriers for the anyons. Finally, the effects of both forms of disorder are studied for coherent transport of the anyons. It is found that the random lattices cause the anyons to move diffusively rather than ballistically, increasing the lifetime of the memory. Adding randomness in the potentials then causes the anyons to localize, leading to further stability.

\acknowledgments

We would like to thank S. Chesi, K. A. van Hoogdalem, and D. P. DiVincenzo for fruitful discussions. This work was partially supported by the Swiss NSF, NCCR Nanoscience, NCCR QSIT, and DARPA.

\begin{appendix}

\section{Critical fraction of random lattices in contact with an Ohmic bath}\label{app:random lattices ohmic}

\begin{figure}[!tb]
\centering
\includegraphics[width=0.95\columnwidth]{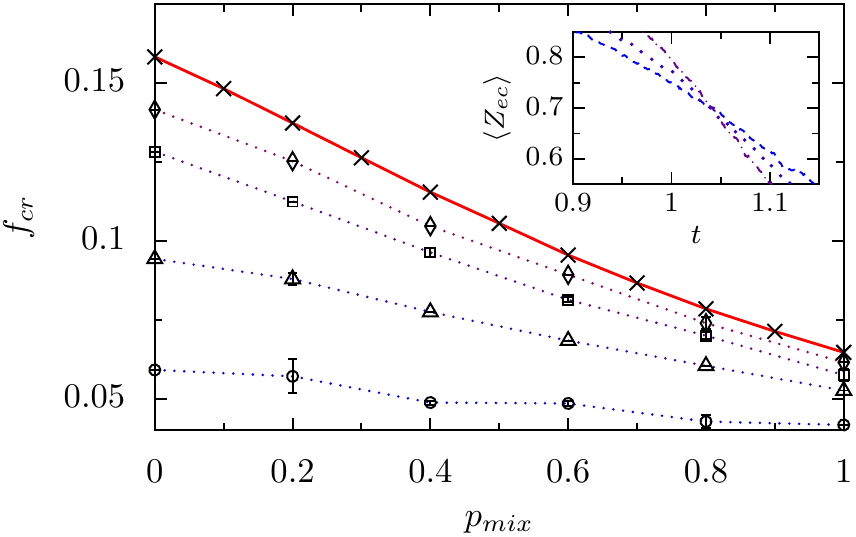}
\caption{(color online) Critical fraction of errors $f_{cr} = f(t=\tau)$ as a function of lattice mixing $p_{mix}$ at temperatures $T = 0.2J$ (circles), $0.45J$ (triangles), $1.0125J$ (squares), and $2J$ (diamonds). The lifetime $\tau$ is given by the intersection of error-corrected logical $Z$ operators for lattice sizes $L = 38, 56, 86$. Error bars are due to the uncertainty in $\tau$. The solid line is determined by the thresholds from the Monte Carlo simulations of an independent error model (see main text). The inset shows an example of crossing logical $Z$ operators for the particular values $p_{mix} = 0.4$ and $T = 0.45J$.}\label{fig:randLatt_ohmic}
\end{figure}

We present here some additional results for the error thresholds of random lattices in the presence of thermal errors. Fig.~\ref{fig:randLatt_ohmic} shows the fraction of errors $f$ at the lifetime of an infinitely large system as a function of the lattice mixing $p_{mix}$ and for different temperatures $T$. In this section, the energy scale is set by the anyon gap $J$. Time is thus measured in units of $(\kappa_1 J)^{-1}$. For given $p_{mix}$ and temperature $T$, we first simulate systems of several different sizes in contact with an Ohmic bath. We then determine the lifetime $\tau$ as the intersection of the decay curves of the corresponding error-corrected logical $Z$ operators (see inset of Fig.~\ref{fig:randLatt_ohmic}). Since the anyon dynamics are independent of the system size (note that the anyons are not interacting with each other), all curves $f(t)$ for different system sizes collapse and the specific value $f(t = \tau) = f_{cr}$ can be read off easily. One can see nicely that these thresholds converge with increasing temperature to the ones given by the model of independent errors. This can be explained by the loss of correlations between errors due to an increasing amount of thermal noise in the form of fluctuating anyons.

\section{Gaussian noise and $1/r$ interaction}\label{app:random gauss}

\begin{figure}[!h]
\centering
\includegraphics[width=0.95\columnwidth]{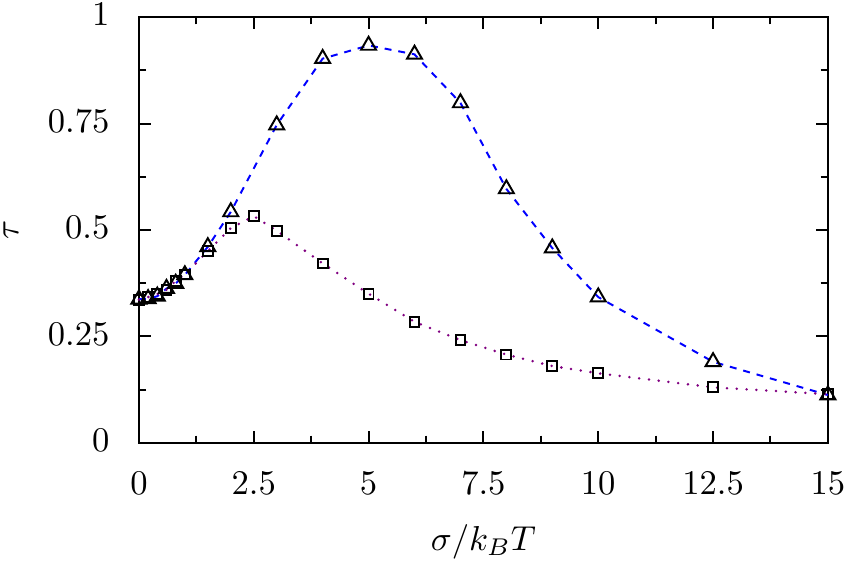}
\caption{(color online) Lifetime $\tau$ as a function of disorder strength $\sigma$ in the presence of Gaussian (triangles) and Ising (squares) noise in an interacting system with $\alpha = 1$, $A = 0.5\, k_B T$. The lines are guides to the eye. The size of this particular system was $L = 32$.}\label{fig:randPot_a1}
\end{figure}

We have also performed simulations with $1/r$ interaction ($\alpha$ = 1) and plot the results in Fig.~\ref{fig:randPot_a1}. Apart from Ising-like disorder ($J_p = \pm \sigma$) we have also looked at a Gaussian distribution of onsite potentials $J_p$ with mean zero and standard deviation $\sigma$. Generally, the lifetimes are shorter than for constant interaction because the weaker $1/r$ interaction allows for a larger density of anyons. Nevertheless, the results are qualitatively similar to the ones discussed in the main text, namely showing a pronounced maxima of the lifetime as a function of $\sigma$. This supports the picture that the interaction is required to limit the number of diffusing anyons, while the initial increase in lifetime with $\sigma$ is due to their obscured diffusion. In the case of Gaussian noise, the latter effect is even stronger, because anyons are created or get trapped in a few sites with onsite potentials much lower than $-\sigma$, out of which it is difficult for them to escape again. This explains the increased lifetime from Ising to Gaussian randomness.

\section{Supporting simulations}\label{app:supporting}

\subsection{Polarized Ising randomness}

In order to confirm the picture that it is indeed the sites with $J_p > 0$ that restrict the diffusion by acting as barriers to the anyons, we have determined the lifetime $\tau$ of an interacting system ($L = 50$, $\alpha = 0$, $A =0.5\, k_B T$, $\sigma = 5\, k_B T$) as a function of the Ising polarization $P$. The latter is defined as $P = 1 - 2\eta$, where $\eta$ is the fraction of sites with negative onsite energy. 

Starting from $P = -1$, i.e., $J_p = -\sigma$ for all sites $p$, the lifetime moderately increases as more and more positive sites are randomly added (increasing $P$). Around $P = 0$, where there is an equal number of sites with positive and negative onsite energies, the lifetime drastically increases by about 2 orders of magnitude. At this point, large connected areas of sites with $J_p < 0$ can no longer exist, such that the anyons can move freely only within areas each consisting of just a few negative sites. Consequently, the diffusion is drastically reduced. If and how the polarization at this threshold is related to the site and bond percolation thresholds of the square lattice, which are $\eta \approx 0.5927$ and $0.5$ (see, e.g., Ref.~\cite{Feng2008}), respectively, is not completely clear.

\begin{figure}[!tb]
\centering
\includegraphics[width=0.95\columnwidth]{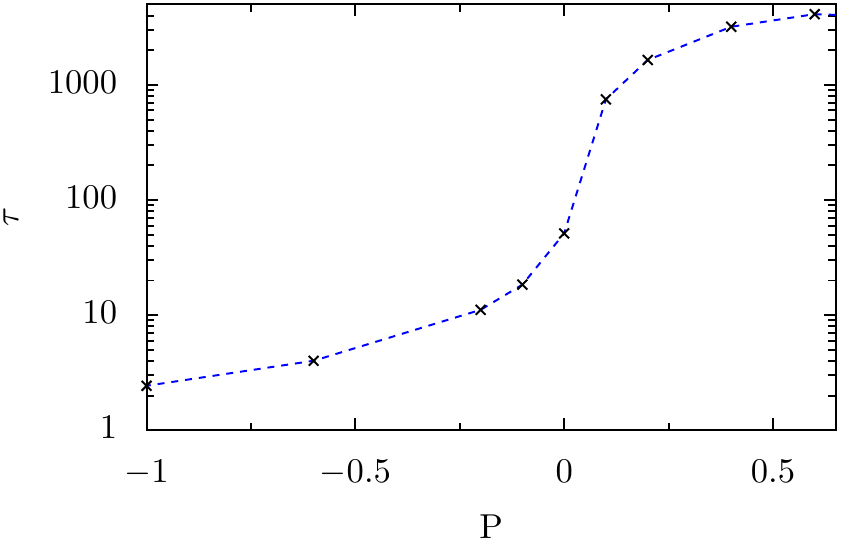}
\caption{(color online) Lifetime as a function of Ising polarization $P$. The parameters for these systems are $L = 50$, $\alpha = 0$, $A = 0.5\, k_B T$, $\sigma = 5\, k_B T$. The dashed line is a guide to the eye.}\label{fig:ising_polarized}
\end{figure}

\subsection{Artificial cutoff of number of anyons}

\begin{figure}[!tb]
\centering
\includegraphics[width=0.95\columnwidth]{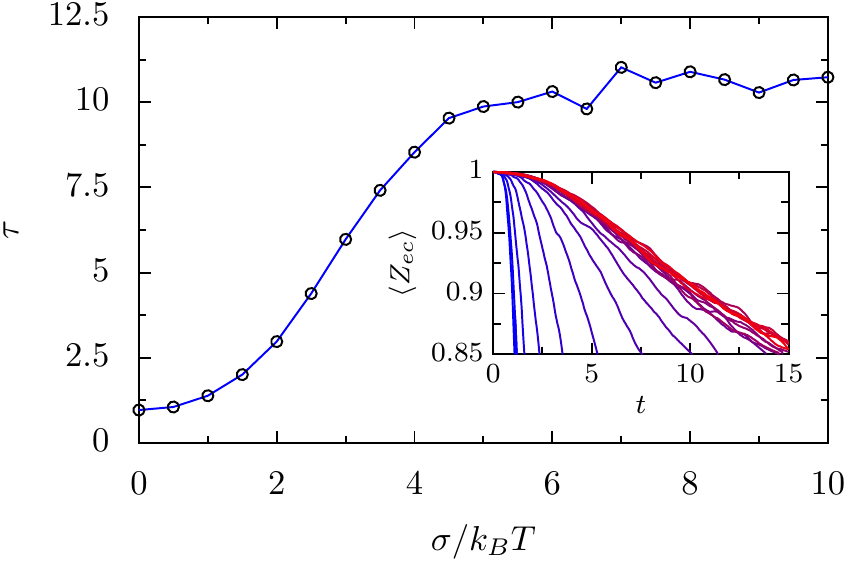}
\caption{(color online) Lifetime as a function of Ising disorder strength in a non-interacting system of size $L = 32$ with an artificially engineered maximal number of anyons equal to $20$. The solid line is a guide to the eye. Inset: The error-corrected logical $Z$ operator as a function of time for different $\sigma$ yielding the lifetimes shown in the main plot.}\label{fig:ising_nonint_cutoff}
\end{figure}

We can support the claim that the only relevant effect of the repulsive interaction is to reduce the number of anyons by simulating a non-interacting system with an artificial maximal number of anyons. This data is shown in Fig.~\ref{fig:ising_nonint_cutoff}. Despite the absence of interaction, the lifetime of encoded states is still growing with increasing Ising disorder strength, hence clearly demonstrating that this effect is caused solely by the disorder. Furthermore, the lifetime saturates for large $\sigma$, because the energy barriers posed by the sites with $J_p = +\sigma$ are essentially infinitely high and increasing them further bears no more advantage. The observed saturation also confirms that it is indeed the growing number of anyons that is responsible for the subsequent decrease in lifetime at large $\sigma$ in the data presented in the main text (where the number of anyons was not artificially restricted).

\end{appendix}

\newpage


%

\end{document}